\documentclass{appolb}
\usepackage{epsfig}

\begin{document}
\title{Presentations from DIS2002 in Krakow}%
\author{Martin McDermott
\address{Division of Theoretical Physics, Dept. Math. Sciences \\
University of Liverpool, L69 3BX, UK}}
\maketitle
\begin{abstract}
Both of my talks at DIS2002, on Generalised Parton Distributions and Nuclear
Shadowing are presented. In an appendix I summarise some of the discussions which followed the talks. 
\end{abstract}
\medskip
  
\section{Introduction}
This document includes both of the talks I gave at DIS2002 in Krakow in May 2002. Section \ref{sec:gpd} contains the proceedings contribution for my talk on 
Generalised Parton Distributions\footnote{In collaboration with Andreas Freund (Regensburg University)}, given in the structure function working 
group. This presentation is closely related to that of Andreas Freund \cite{afdis2002} concerning our NLO QCD analysis of Deeply Virtual Compton Scattering \cite{afmm, website}
in the diffractive working group. Ruben Sandapen also gave a talk \cite{rsdis2002} on this 
process in the same session, looked at from the dipole model framework, which presented results from our joint paper \cite{mss}.
Section \ref{sec:npdfs} contains my proceedings contribution on nuclear shadowing and diffraction, given in the diffractive interactions working group and based on the analysis of \cite{fgms}. The transparencies for both talks may be found on my homepage \cite{home}.
Finally, in the appendix of section \ref{sec:app}, I comment on some of the issues raised at the meeting pertaining to these talks.   

\section{Generalized parton distributions at next-to-leading order}

\label{sec:gpd}

{\it Abstract: This talk discusses generalized parton distributions (GPDs), 
which encode various types of non-perturbative information
relevant to the QCD description of exclusive processes. 
Results on their next-to-leading order (NLO) QCD evolution are presented.  
We find that models for the input GPDs based on double distributions 
require some modification in order to reproduce the available data on 
deeply virtual Compton scattering.}

\begin{figure} 
\centering 
\mbox{\epsfig{file=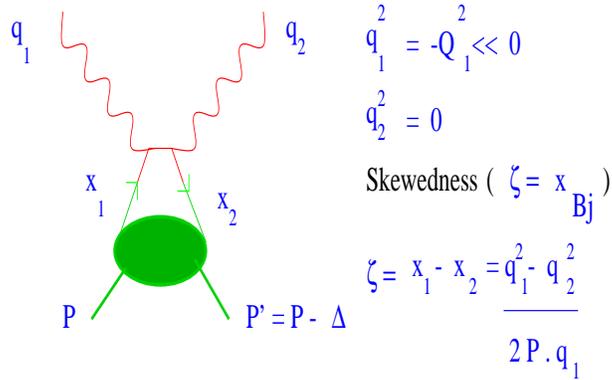,width=8cm,height=5cm}} 
\caption{DVCS amplitude and skewed kinematics} 
\label{fig1} 
\end{figure}

Generalized parton distributions (GPDs) are required to calculate a wide variety 
of hard exclusive processes (\eg diffractive electroproduction of vector mesons, 
or dijet photoproduction). The easiest and cleanest way to access GPDs is via 
the electroproduction of a real photon, \ie Deeply Virtual Compton Scattering 
\cite{afdis2002} (see figure \ref{fig1}, which also defines some kinematic variables). 
DVCS amplitudes have been proven to factorize \cite{jcaf}, \ie to involve 
convolutions of perturbatively calculable coefficient functions with GPDs. 
The ZEUS, H1, HERMES and CLAS experiments all have data available \cite{data}. 
On the theoretical side the next-to-leading order (NLO) leading-twist analysis 
of DVCS is now complete and a great deal has been understood about the role of 
higher twist corrections (see \eg \cite{bm} and references therein). We have completed a 
NLO numerical analysis of GPDs, DVCS amplitudes and observables \cite{afmm} 
and present some of our results here. Most of our analysis code is 
available from the HEPDATA website \cite{website}.

\begin{figure} 
\centering 
\mbox{\epsfig{file=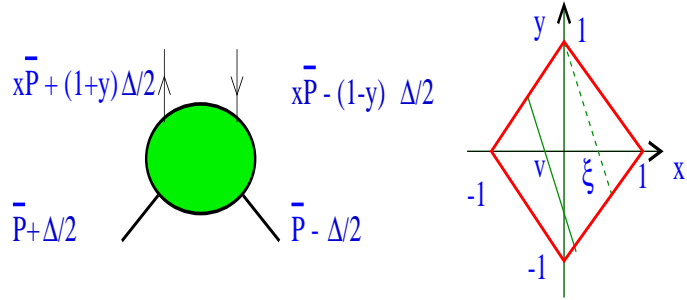,width=9cm,height=4cm}} 
\caption{Symmetric double distributions and their physical domain} 
\label{fig2} 
\end{figure}

GPDs are defined by Fourier transforms of twist two operators sandwiched between unequal 
momentum nucleon states. They encode a variety of non-perturbative information about the 
nucleon, including conventional parton distribution functions (PDFs), distributions 
amplitudes and form factors, and reproduce these in various limits. The essential feature 
of the two parton correlation function shown in figure \ref{fig2} is the presence of a finite  
momentum transfer, $\Delta = P-P'$, in the $t$-channel. Hence the partonic structure of the hadron is tested at {\it distinct} momentum fractions, $x_1, x_2$. On the light cone these matrix elements are parameterized by double distributions (DDs) which depend on two plus-momentum fractions with respect to two external momenta, on the four momentum transfer squared, $t = \Delta^2$, and on a four-momentum scale $\mu^2$. The external momenta can be selected in several 
ways (\eg either the ``symmetric'' ($\Delta, {\bar P} = (P+P')/2 $), or ``natural'' ($\Delta, P$) choices). 
Unfortunately this freedom has led to a proliferation of definitions and 
nomenclature in the literature (skewed, off-diagonal, non-diagonal, off-forward, $\cdots$) to 
describe essentially the same objects, which has led to considerable confusion. 
Hence the collective name {\it generalized} has been introduced to attempt to clarify 
the situation.

\begin{figure} 
\centering 
\mbox{\epsfig{file=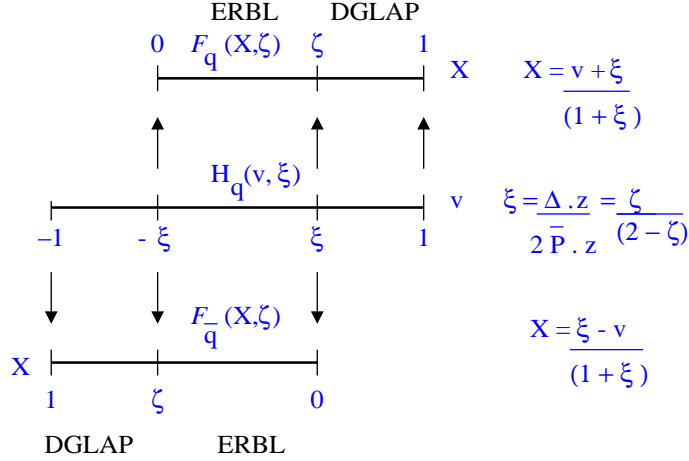,width=9.0cm,height=6cm}} 
\caption{Relation between ${\cal F}$ and $H$} 
\label{fig3} 
\end{figure}

Radyushkin \cite{rad1} introduced symmetric DDs, with plus momentum fractions, $x,y$ of the outgoing and returning partons defined as shown in the left hand plot of figure \ref{fig2}. They exist on the diamond-shaped domain shown to the right. For a given skewedness, $\xi = \zeta/(2-\zeta)$, the outgoing parton lines of course only have a single plus 
momentum, so that Ji's distributions $H(v,\xi)$ \cite{ji} are related to these DDs, 
via an integral involving $\delta(v - x - \xi y)$,  along the off-vertical lines in the 
diamond ($v \in [-1,1]$, and the dotted line corresponds to $v=\xi$). For our
numerical solution of the renormalization group equations we prefer to work with the 
{\it natural} off-diagonal PDFs defined by Golec-Biernat and Martin \cite{kgbm}, which have a momentum fraction $X \in [0,1]$ of the incoming proton's momentum. Their relationship to 
Ji's functions is shown in figure \ref{fig3}. There are two distinct regions: the DGLAP 
region, $X > \zeta$, in which the GPDs obey a generalized form of the DGLAP equations for PDFs, and the ERBL region, $X<\zeta$, where the GPDs obey a generalized form of the ERBL equations for distributions amplitudes. In the ERBL region, due to the fermion symmetry, ${\cal F}_q$ and ${\cal F}_{\bar q}$ are not independent and this leads to an anti-symmetry of the unpolarised quark distributions about the point $\zeta/2$ (the gluon GPD is symmetric). 
Another formal property of the GPDs, which can be proved on general grounds, is that the N-moments of $H$ are polynomials of degree $\xi^{N}$: this is known as polynomiality. In addition, any input model for GPDs must reproduce the conventional PDFs for very small skewedness: 
$\stackrel{\mathrm{lim}}{_{\zeta \rightarrow 0}} {\cal F}_i (X,\zeta) \rightarrow f_i (X)$: the ``forward limit''.

As a model for the GPDs, with the correct features, we use Radyushkin's factorized ansatz \cite{rad1} for the double distributions:

\begin{equation}
F^{DD} (x,y) = \pi (x,y) f^i (x) A^{i} (t) 
\end{equation}

\noindent where $A^i (t)$ is a form factor form for the factorized $t$-dependence, $f^i (x) $ is the forward PDF and 
\begin{equation}
\pi (x,y) = \frac{\Gamma (2b + 1)}{2^{2b+1} \Gamma^2 (b + 1)} \frac{[(1-|x|)^2 - y^2]^b}{(1-|x|)^{2b+1}} 
\end{equation}
is the profile function which introduces the dependence on skewedness (normalised such that $\int^{1-|x|}_{-1+|x|} dy ~\pi (x,y)  = 1$).  
In the canonical model $b_q=1$ and $b_g=2$, $b=\infty$ corresponds to the forward case.  
By design this model automatically respects the forward limit. To respect polynomiality an additional term, the so-called D-term, is required in the ERBL region, for which we use the model of \cite{polyweiss}. Numerical studies indicate that this term is significant only at large $\zeta$ (its influence drops below $1\%$ in for $\zeta < 0.01$).

In the DGLAP region integration over $y$ of the DD leads to the follow type of integral:
\begin{equation}
{\cal F}_{q,a} (X,\zeta) = \frac{2}{\zeta} \int^{X}_{\frac{X-\zeta}{1-\zeta}} dx' \pi^q (x', \frac{v-x'}{\xi}) q^{a} (x') \, .
\end{equation}
\noindent This leads to a serious problem in this model: when $X \rightarrow \zeta$ the PDF is sampled down to zero, where it has not yet been measured. 
For non-singular distributions this presents no particular problems (although it does involve an extrapolation to $x'=0$), however, for 
singular distributions the precise extrapolation is crucial and in general leads to a large enhancement of the GPD relative to the PDF 
in this region (for CTEQ6M the factor can be as large as five). When we compared this 
model to the H1 DVCS data it overshoots by a factor of approximately 4-6 because of this ! 
To tame this rather unnatural enhancement we introduce a modification of such integrals, 
via a lower cutoff of the form $a ~\zeta$. This may be justified by examining 
the effect of imposing exact kinematics on the imaginary part of the DVCS amplitude which 
would be required to produce finite mass hadrons in the final state. Such reasoning 
indicates that $a \sim m^2_{\mathrm{hadron}}/Q_0^2 \approx 1/2$ is a 
reasonable value.
Introducing such a cutoff reduces the enhancement factor of the input GPD close to $X=\zeta$ considerably 
and allows the H1 data to be well described at both LO and NLO. Unfortunately, 
it leads to a mild violation of the polynomiality condition since it may introduce higher moments, or slightly alter the highest allowed moments.

\begin{figure} 
\centering 
\mbox{\epsfig{file=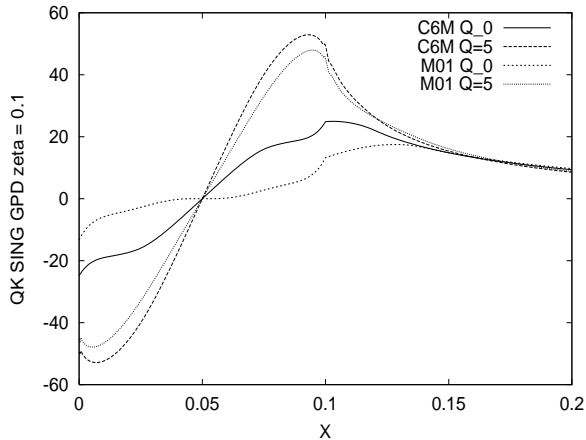,width=8cm,height=6cm}} 
\caption{Quark singlet GPD for $\zeta = 0.1$ at the input scale $Q_0$ and $Q=5$}
\label{fig4} 
\end{figure} 

Both the continuity of the GPD through the boundary point, $X=\zeta$, and the symmetries about the point 
$X=\zeta/2$ are preserved under evolution. The evolution equations, at NLO accuracy, are 
solved numerically on a grid for each value of $\zeta$. For example figure \ref{fig4} shows the quark GPD at $\zeta = 0.1$ at the input scale 
and evolved to $Q=5$~GeV for CTEQ6M (C6M) and MRST01 (M01) \cite{pdfs} input PDFs. This figure demonstrates that the anti-symmetry about $\zeta/2$ is preserved under evolution.

The H1 data is already starting to constrain the allowed input GPDs.
At present input models are based only on the formal mathematical properties of the GPDs 
(polynomiality, symmetries and the forward limit). 
As the data improves it will become necessary to fit the input distributions via minimization methods in a 
similar fashion to the inclusive case. Our analysis indicates that the cutoff parameter, $a$, and the profile 
function power, $b$, may be good candidates for fit parameters, since both of them control the level of skewedness imposed at the input scale.

\section{Diffraction and nuclear shadowing}

\label{sec:npdfs}

{\it Abstract: I present results from our recent leading twist QCD analysis of nuclear 
shadowing\footnote{In affiliation with V. Guzey (Adelaide University), L. Frankfurt (Tel Aviv University) and  M. Strikman (Penn State University)} and contrast them with predictions using the eikonal model. 
By exploiting QCD factorization theorems, the leading twist approach 
employs diffractive parton distributions, extracted from diffractive DIS
measurements at HERA, to calculate the nuclear shadowing correction on the 
parton level. Large nuclear shadowing effects are found for the gluon channel which are reflected in the predictions for $F_L^A$.}

The most naive assumption about deep inelastic scattering (DIS) on a nucleus 
one can make is that the photon scatters independently of each nucleon, 
which gives for the nuclear structure function: $F_2^A = A F_2^N$. 
For small $x \mathrel{\rlap{\lower4pt\hbox{\hskip1pt$\sim$}}
    \raise1pt\hbox{$<$}} 0.05$ the main negative nuclear correction is nuclear shadowing, 
\ie the coherent interaction of the photon with several nucleons at once, 
which leads to $F_2^A/ A F_2^N < 1$. For a low density nucleus, nuclear shadowing 
is closely related to diffraction off a nucleon. The leading twist QCD analysis of 
Frankfurt and Strikman \cite{fs} relates nuclear PDFs to diffractive parton 
distribution functions (DPDFs), by exploiting the QCD factorization theorem 
for inclusive diffraction \cite{col}. In this talk I present some results from the 
detailed analysis of \cite{fgms} which exploited the latest available DPDFs 
to make predictions for nuclear shadowing and hence nuclear PDFs.
The leading twist approach has a sharply contrasting space-time picture and 
predictions to the popular eikonal approach to nuclear shadowing (which is 
closely related to the $q {\bar q}$-dipole model of diffraction).

Why should one be interested in DIS on a nucleus ? Firstly, nuclear PDFs 
provide boundary conditions for novel process (\eg the search for beyond the 
standard model effects, or for new matter states in QCD such as the quark 
gluon plasma) as well as for ``standard'' QCD processes in nuclear collisions. 
Secondly, a particularly interesting feature of eA is the access to a high parton 
density regime at much lower energies than in DIS on nucleons. Thirdly, 
because of the intimate connection between nuclear shadowing and 
diffraction on a nucleon, high statistics data on nuclear PDFs could be used 
to discriminate between competing models of diffraction. Lastly, the study of 
DIS on nuclei is timely since high energy electron nucleus collisions are 
currently being considered seriously for HERA after 2006 and there is an 
electron-ion collider (EIC) planned for the USA circa 2012.

\begin{figure} 
\centering 
\mbox{\epsfig{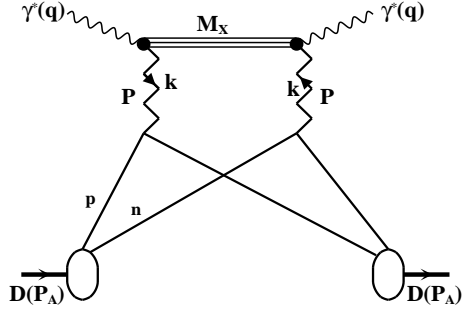}} 
\caption{Nuclear shadowing diagram in virtual photon deuteron scattering.} 
\label{fig5} 
\end{figure}

The starting point of the leading twist approach to nuclear shadowing is the 
application of the logic of Gribov \cite{gribov} to DIS on the deuteron. 
The optical theorem relates the total cross section, $\sigma_{\mbox{tot}} (\gamma^* D)$ 
to the imaginary part of the forward scattering amplitude. On the forward 
amplitude level the photon may interact elastically with either the proton 
or the neutron or diffractively with both. The latter case corresponds to the 
nuclear shadowing correction (see figure \ref{fig5}). Hence the nuclear 
shadowing correction, $\delta F_{2}^{D}=F_{2}^{p}+F_{2}^{n}-F_{2}^{D}$, 
can be expressed in terms of the structure function for the diffractive scattering 
of the photon off a nucleon:
\begin{equation}
\delta F_{2}^{D}(x,Q^2)=2\frac{1-\eta^2}{1+\eta^2}\int dt dx_{I\!\!P} F_{2}^{D(4)}(\beta,Q^2,x_{I\!\!P},t) F_D(4t+4x_{I\!\!P}^2m_N^2) 
\end{equation}
\noindent where the prefactor, involving $\eta = Re A^D / Im A^D = \pi/2 (\alpha_{I\!\!P}(0) -1)$ \, comes from the AGK cutting rules and $F_D$ is the deuteron form factor.

The result for deuteron generalises easily to any pair of nucleons in a nucleus with $A$ nucleons:
\begin{eqnarray}
&\delta F_{2}^{A(2)} = \frac{A(A-1)}{2} 16 \pi Re \Bigg[ \frac{(1-i\eta)^2}{1+\eta^2} \int d^2b \int_{-\infty}^{\infty} dz_1 \int_{z_{1}}^{\infty} dz_2 \int_x^{x_{I\!\!P,0}} dx_{I\!\!P} \times \nonumber \\
&F_{2}^{D(4)}(\beta, Q^2, x_{I\!\!P}, k_t^2) \bigg|_{k_t^2=0} \rho_A(b,z_1)\rho_A(b,z_2) e^{ix_{I\!\!P}m_N(z_1-z_2)} \Bigg] 
\end{eqnarray}
\noindent where $\rho $ is the nucleon density, $z, b$ are longitudinal position and impact parameter of the nucleon concerned. The interaction with more than two nucleons requires some modelling, and we invoke $\sigma_{eff}$ for the rescattering cross section (calculated in the quasi-eikonal approximation).

\begin{figure} 
\centering 
\mbox{\epsfig{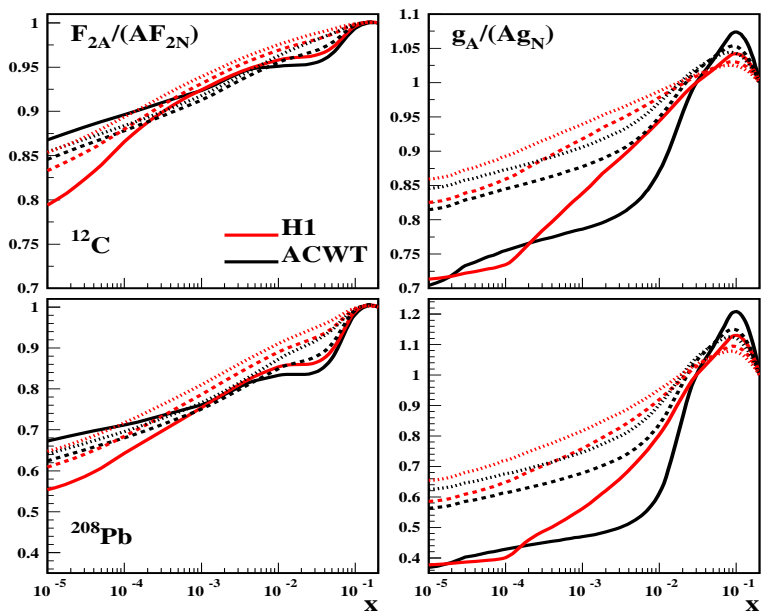}} 
\caption{Predictions for $ F^{A}_{2}$ and nuclear gluon PDFs for $Q=2,5,10$~GeV (solid, dashed, dotted curves). In each case two curves are shown to indicate the spread associated with extreme choice of DPDFs models.}
\label{fig6} 
\end{figure}

Since inclusive and diffractive structure functions both factorize, 
and have the same coefficient functions, one can factor off
the hard pieces to relate the PDFs themselves. Hence, on the parton level
\begin{eqnarray} 
&\delta f_{j/A}(x,Q^2)  = \frac{A(A-1)}{2} 16 \pi Re \Bigg[\frac{(1-i\eta)^2}{1+\eta^2} 
\int d^2b \int_{-\infty}^{\infty} dz_1 \int_{z_{1}}^{\infty} dz_2 \int_x^{x_{I\!\!P, 0}} 
dx_{I\!\!P} \times  \\
& f_{j/N}^{D}(\beta, Q^2,x_{I\!\!P},0) \rho_A (b,z_1) \rho_A (b, z_2) e^{ix_{I\!\!P} 
m_N(z_1-z_2)} e^{-(A/2) (1-i\eta) \sigma_{eff}^{j} \int^{z_{2}}_{z_{1}} dz \rho_{A}(z)} \Bigg] \nonumber
\end{eqnarray}
\noindent where $f_{j/N}^{D} $ is the DPDF for a parton of flavour $j$. The exponential factor in the last line calculates the rescattering. We used several models of DPDFs, tuned to the HERA data \cite{dmodels}. 
For the effective cross section for rescattering of octet configurations we found that it was necessary to introduce corrections to prevent unitarity being violated 
(so called saturation effects). Since gluon DPDFS are large we find a corresponding 
large nuclear shadowing for gluons (see figure \ref{fig6}). 
For a discussion of enhanced nuclear shadowing correction for central collisions, 
of the uncertainties associated with the unmeasured diffractive slope, and of the 
implementation of charm, we refer the reader to our paper \cite{fgms}.

\begin{figure} 
\centering 
\mbox{\epsfig{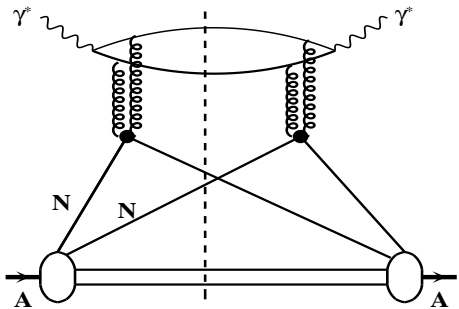}} 
\caption{The eikonal model for nuclear shadowing.}
\label{fig7} 
\end{figure}

\begin{figure} 
\centering 
\mbox{\epsfig{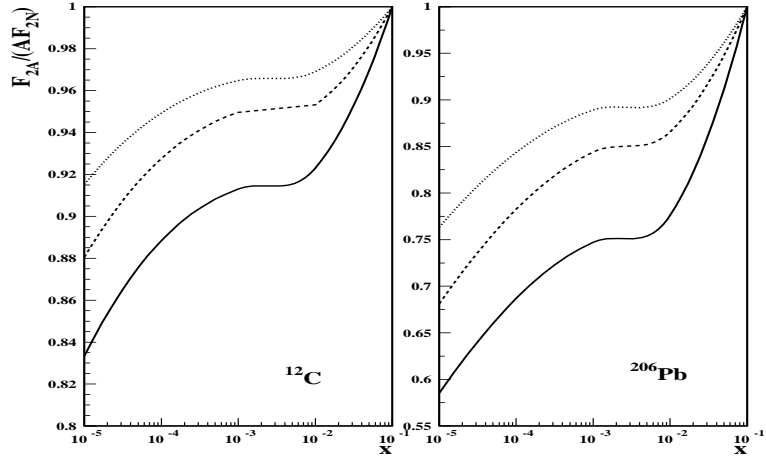}} 
\caption{Eikonal model predictions for $F_2^A$ for $Q=2,5,10$~GeV (solid, dashed, dotted curves).}
\label{fig8} 
\end{figure}

\begin{figure} 
\centering 
\mbox{\epsfig{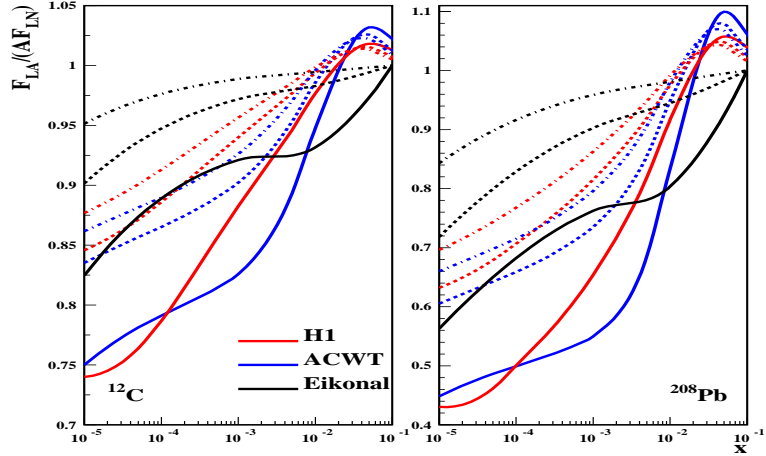}} 
\caption{Contrasting predictions from the eikonal model (black) and the leading twist model (blue and red) for $F_L^A$ for $ Q=2,5,10$~GeV (solid, dashed, dotted curves).}
\label{fig9} 
\end{figure}

In the eikonal model for nuclear shadowing \cite{eik} the $q {\bar q}$ pair 
scatters elastically off many nucleons in the target (see figure \ref{fig7}). 
The fundamental interaction of the dipole with a nucleon is eikonalized and 
the formula is given by
\begin{eqnarray}
\delta F_{2}^{A}(x,Q^2) = &\frac{Q^2}{4 \pi^2 \alpha_{em}} \frac{A(A-1)}{2} Re \Bigg[ (1-i\eta)^2 \int d\alpha ~d^2 d_{t} \sum_{i}~|\psi(\alpha,Q^2,d_{t}^2)|^2 \times \nonumber \\ 
& \int d^2b \int_{-\infty}^{\infty} dz_1 \int_{z_{1}}^{\infty} dz_2 \Big[\sigma^{tot}_{q\bar q N}(x,d_{\perp}^2,m_{i})\Big]^2 ~\rho_A(b,z_1) \rho_A(b,z_2) \times \nonumber \\
&\, e^{i2x m_N(z_1-z_2)} e^{-(A/2)(1-i\eta) \sigma^{tot}_{q\bar q N}(x,d_{\perp}^2,m_{i}) \int^{z_{2}}_{z_{1}} dz \rho_{A}(z)} \Bigg] 
\end{eqnarray}
\noindent where $\psi$ is the light-cone wavefunction for 
$\gamma^* \rightarrow q {\bar q} $, taken from QED ($\alpha$ is the momentum 
fraction carried by the quark). Generally, the mixing with higher Fock states 
in the virtual photon is neglected (this implies that $Q^2$-dependence is not 
consistent with DGLAP !). 
Hence a parton level description of  nuclear shadowing in the eikonal 
model is impossible. To implement the eikonal model we employed the MFGS-dipole 
\cite{mfgs} for $\sigma^{tot}_{q\bar q N}$, but we could also have used other dipole models. 
In the eikonal model nuclear 
shadowing is suppressed by colour transparency (since  $\sigma \propto d_t^2$, at 
small transverse size $d_t$). This implies that nuclear shadowing of $F^A_{2}$ decreases 
rapidly with increasing $\, Q$ (see figure \ref{fig8}). This higher-twist nature of 
nuclear shadowing in the eikonal model is clearest for hard processes, which are most sensitive to small size configurations (\eg  $F^{A}_{L}$ at large $Q^2$, 
see figure \ref{fig9}).

To conclude, the leading twist QCD analysis of \cite{fgms} suggests that nuclear 
shadowing is a leading twist phenomena. It produces radically different predictions to 
the eikonal approach popular in the literature, (cf. the dipole model for diffraction) 
for which nuclear shadowing is a higher twist effect. A systematic measurement of nuclear
 PDFs (via $F_{2, L}^{A}$, nuclear DVCS and Drell-Yan, which should be possible at 
HERA III and EIC), and hence of nuclear shadowing, can help establish the correct 
model for diffraction. 
It may well be possible to investigate non-linear QCD in DIS on a large nucleus, but
one needs to understand nuclear shadowing first.

\section{Appendix: discussions}

\label{sec:app}

In the discussion session the question was raised whether the 
skewedness effect necessarily leads to an enhancement in general at NLO 
(as is the case at LO). 
Unfortunately, it is not possible to give a complete answer at this time. 
Within the framework of the double distribution input model one sees a 
suppression of the GPD at the input scale for a falling forward distribution as $x \rightarrow 0$, and an enhancement for a rising distribution, 
relative to the forward case.
It is clear however that skewed evolution produces an enhancement relative 
to forward evolution of a given input. So, even for the relatively clean 
DVCS process, whether one sees an enhancement or a suppression depends on 
what one chooses for the input GPD and how close one is to the input scale.

As J. Bartels pointed out correctly, to answer this question for a particular 
process, for example diffractive vector meson production of $J/\psi$, it is 
necessary to have a complete NLO calculation, which includes calculating 
the $q {\bar q} g$ component of the virtual photon and diffractively 
produced system. At present the NLO coefficient function are known only 
for DVCS \cite{bm}. It is clear that a full NLO analysis is required for 
each process. 

A global analysis of high energy exclusive processes, \ie DVCS and 
photoproduction and electroproduction of all vector meson states, 
within the framework of the dipole model, is also highly desirable. 
In order to proceed in a quantitative fashion, it would be optimal 
to place the semi-qualitative dipole model developed in \cite{mfgs} 
on a firmer footing by performing a fit to the available inclusive 
structure function data. Such a fit is currently under active study.
One could then use this ``fitted'' dipole cross section in the global 
analysis of exclusive processes (remembering to take skewedness into 
account, as appropriate).

Following my talk on nuclear parton distributions a question was raised 
concerning a comparison to the available data. My collaborator V. Guzey 
is currently looking into this issue. He also plans to provide analytic 
forms for the nuclear parton distribution functions fitted to the numerical 
results obtained in our paper \cite{fgms}.

\section*{Acknowledgments}

I'd like to thank the organisers for an excellent conference in the beautiful setting of Krakow. It is a pleasure to thank my colleagues for interesting and enjoyable discussions, particularly those in the ``evening sessions'' in the old town !

\end{document}